\documentclass[12 pt]{JHEP3}
\usepackage[centertags]{amsmath}
\usepackage{amsfonts}
\usepackage{amssymb}
\usepackage{amsthm}
\usepackage{newlfont}
\usepackage{tensind}\tensordelimiter{?}
\usepackage{graphicx}

\renewcommand{\thesubsection}{\arabic{section}.\arabic{subsection}}

\newcommand{\eps}{\varepsilon}

\newcommand{\Rf}{\mathbb{R}}
\newcommand{\Zf}{\mathbb{Z}}\newcommand{\Nf}{\mathbb{N}}

\newcommand{\dd}{\partial}

\newcommand{\dif}[2]{\frac{d #1}{d #2}}

 \DeclareMathOperator{\tr}{Tr}

\newcommand{\tbyt}[4]{\begin{pmatrix} #1 & #2 \\ #3 & #4 \end{pmatrix}}
\newcommand{\vt}[2]{\begin{pmatrix} #1 \\ #2\end{pmatrix}}

\newcommand{\ra}{\rightarrow}

\newcommand{\x}[1]{\text{#1}}

\newcommand{\ads}{AdS$_2$ }
\author{Ofer Aharony and Assaf Patir\\\\
Department of Particle Physics,\\
Weizmann Institute of Science,\\ Rehovot 76100, Israel\\ E-mail:
\email{Ofer.Aharony@weizmann.ac.il},
\email{Assaf.Patir@weizmann.ac.il}} \abstract{We analyze the
conformal limit of the matrix model describing flux backgrounds of
two dimensional type 0A string theory. This limit is believed to be
dual to an \ads background of type 0A string theory.  We show that
the spectrum of this limit is identical to that of a free fermion on
AdS$_2$, suggesting that there are no closed string excitations in
this background.} \preprint{WIS/22/05-SEP-DPP} \title{The Conformal
Limit of the 0A Matrix Model and String Theory on AdS$_2$}
\begin{document}
\section{Introduction and Summary}
Two dimensional non-critical string theories are useful toy models
for studying various aspects of string theory (for reviews see
\cite{Klebanov:1991qa,Ginsparg:1993is,Polchinski:1994mb}).
In particular, the two dimensional type 0 string theories are useful
for this, since they are non-perturbatively stable and they have a
known matrix model dual \cite{Douglas:2003up,Takayanagi:2003sm}.
Since the type 0 theories have Ramond-Ramond (RR) fields, they can
be used to study RR backgrounds, whose worldsheet description in
string theory is still poorly understood.

The type 0A theory has RR 2-form field strengths which are sourced
by D0-branes. It has been suggested that this theory should have an
\ads solution with RR flux. This is based both on the fact that the
space-time effective action has an extremal black hole solution
whose near-horizon limit is AdS$_2$, and on the fact that the matrix
model for the 0A theory with flux has a limit in which it is a
conformally invariant quantum mechanical system (which one expects
to be dual via the AdS/CFT correspondence to an \ads background of
string theory) \cite{Gukov:2003yp,Strominger:2003tm}. If such a
background exists, it is interesting to study it, both as a simple
example of a RR background (the large symmetry may help in finding a
useful worldsheet description of this background) and as an example
of the correspondence between \ads backgrounds and conformal quantum
mechanics, which is still not as well understood as other examples
of the AdS/CFT correspondence.

In this note we study the conformal limit of the 0A matrix model
in order to find the properties of this conjectured \ads
background. We find that the spectrum of this theory is equal to
that of a free fermion field on AdS$_2$, with a mass proportional
to the RR flux $q$. This fermion originates from the eigenvalues
of the matrix model which correspond to D0-anti-D0-brane pairs, so
this spectrum suggests that the only excitations in this \ads
background are such uncharged D-brane-pairs, and that there are no closed
string excitations in this background. From the point of view of
0A backgrounds with flux which asymptote to a linear dilaton
region, this implies that the closed string excitations cannot
penetrate into the strongly coupled region which is dual to the
conformal limit of the matrix model.

We begin in section
\ref{type0} by reviewing two dimensional type 0A string theory,
its matrix model description, and its extremal black hole
background. In section \ref{ads2cqm} we analyze the conformal
limit of the matrix model and compute its spectrum. In section
\ref{strcqm} we discuss the implications of this spectrum for the
dual string theory on AdS$_2$, and in section \ref{boso} we
discuss bosonizations of the fermionic system we find, which may
be useful for studying the conformal quantum mechanics with a
non-zero Fermi level. In the appendix we provide a detailed
computation of the spectrum of a spinor field on AdS$_2$.
%
%
\section{A Brief Review of Type 0A Superstrings}\label{type0}
\subsection{Type 0A Spacetime Effective Action}
Two dimensional fermionic strings are described by $\mathcal N=1$
supersymmetric worldsheet field theories coupled to worldsheet
supergravity. The non chiral GSO projection gives the two type 0
theories. Because there are no NS-R or R-NS sectors in the theory,
the type 0 theories have no fermions and are not supersymmetric in
spacetime. The NS-NS sector includes a graviton, a dilaton and a
tachyon. In two dimensions, there are no transverse string
oscillations, and longitudinal oscillations are unphysical except
at special values of the momentum. Thus, the tachyon is the only
physical NS-NS sector field. In type 0A, the R-R sector
contributes two 1-forms, and the action is \cite{Douglas:2003up}
(with $\alpha'=2$)
\begin{multline}\label{type0action}
    S=\int d^2x\sqrt{-g}\bigg[\frac{e^{-2\Phi}}{2\kappa^2}\bigg(
    4+R+4(\nabla\Phi)^2-\frac12(\nabla
    T)^2+\frac12T^2+\dotsb\bigg)-\\-\pi e^{-2T}(F^{(+)})^2-
    \pi e^{2T}(F^{(-)})^2+\dotsb\bigg].
\end{multline}
The equations of motion are solved by a linear dilaton solution
\begin{align}
    &G_{\mu\nu}=\eta_{\mu\nu},&\Phi=x,
\end{align}
where $x$ is the spatial coordinate. A possible deformation is to
turn on the tachyon (whose mass is lifted to zero by the linear
dilaton) $T=\mu e^x$.

The general equations of motion are
\begin{subequations}
\begin{align}\label{eomg}
\begin{split}
    0&=2R_{\mu\nu}+4\nabla_\mu\nabla_\nu\Phi-\nabla_\mu T\nabla_\nu
    T+2\pi\kappa^2e^{2\Phi-2T}[g_{\mu\nu}(F^{(+)})^2-4
    (F^{(+)})_\mu^\beta
    (F^{(+)})_{\nu\beta}]+\\
    &+2\pi\kappa^2e^{2\Phi+2T}[g_{\mu\nu}(F^{(-)})^2-4 (F^{(-)})_\mu^\beta
    (F^{(-)})_{\nu\beta}],
\end{split}\\
    0&=-2-\nabla^2\Phi+2(\nabla\Phi)^2-\frac14T^2
    -\pi\kappa^2e^{2\Phi}[e^{-2T}(F^{(+)})^2+e^{2T}(F^{(-)})^2],\\
    0&=\nabla^2T-2\nabla\Phi\cdot\nabla T+T
    +4\pi\kappa^2e^{2\Phi}[e^{-2T}(F^{(+)})^2-e^{2T}(F^{(-)})^2],
\\\label{eomF}
    0&=\nabla^\nu(e^{\mp2T}F^{(\pm)}_{\mu\nu}).
\end{align}\end{subequations}
Equation \eqref{eomF} implies
\begin{equation}
    0=\nabla_\nu(e^{\mp2T}F^{(\pm)\mu\nu})=\dd_\nu(\sqrt{-g}e^{\mp2T}F^{(\pm)\mu\nu})
    =\begin{cases}-\dd_0(\sqrt{-g}e^{\mp2T}F^{(\pm)01})\\
    \dd_1(\sqrt{-g}e^{\mp2T}F^{(\pm)01})
    \end{cases},
\label{solF}
\end{equation}
i.e. $\sqrt{-g}e^{\mp2T}F^{(\pm)01}$ is constant. Thus, we see
that the zero modes are the only degrees of freedom of the vector
field. Furthermore, due to the non trivial coupling in front of
$(F^{\pm})^2$, a time independent field strength can only be
turned on for the vector field whose coupling $e^{\pm2T}$ does not
vanish at infinity. For example, in the linear dilaton background,
the solutions of \eqref{solF} are
\begin{align}
     F^{(+)}_{01}&=q^+e^{+2T},
    &F^{(-)}_{01}&=q^-e^{-2T}.
\end{align}
Suppose that $\mu<0$. As $x\ra\infty$, $T\ra-\infty$ and $F^{(-)}$
becomes singular, while $F^{(+)}$ is regular. Thus, turning on
$F^{(-)}$ requires having D-branes as a source for this field,
while no such branes are needed for $F^{(+)}$. These D-branes
carry RR 1-form charge, so these are D0-branes (known as
ZZ-branes). For positive $\mu$ the situation is reversed. Thus,
both $q^+$ and $q^-$ are quantized. For $\mu=0$ it seems that we
can turn on both fields, however, as shown in
\cite{Maldacena:2005he}, this requires the insertion of $q^+q^-$
strings that stretch from $x=-\infty$ to the strongly coupled
region. This leads to additional terms in the effective action
\eqref{type0action} (see also \cite{Maldacena:2005hi}).
\subsection{Matrix Model Description}
In analogy to the bosonic case \cite{McGreevy:2003kb}, it was
conjectured in \cite{Douglas:2003up,Takayanagi:2003sm} that the
field theory on $N$ such D0-branes (with $N\ra\infty$) is a dual
description of the full string theory. It was argued there that
the matrix model that gives the linear dilaton background of the
0A theory is a U$(N)_A\times$U$(N)_B$ gauge theory with a complex
matrix $m$ in the bifundamental representation. The two $U(N)$
groups originate from D0 and anti-D0 branes.

There are two ways of introducing RR flux \cite{Douglas:2003up}.
First, we can modify the
gauge group to U$(N)_A\times$U$(N+q^-)_B$. This leads to $q^-\neq0$,
$q^+=0$ and corresponds to placing $N+q^-$ ZZ-branes and $N$
anti-ZZ-branes at $x=+\infty$. As long as the Fermi level is below
the barrier ($\mu<0$), we will have $q^-$ charged ZZ-branes left
over after the open string tachyon condenses, so this is expected to
correspond to the background with $q^-$ units of $F^{(-)}$ flux and
no $F^{(+)}$ flux. When reducing this rectangular matrix model to
fermionic eigenvalue dynamics one finds that the potential for the
eigenvalues is \cite{difrancesco}
\begin{equation}\label{eigpot}
    V(\lambda)=-\frac18\lambda^2+\frac M{2\lambda^2},
\end{equation}
where $M=(q^-)^2-1/4$ and $\lambda$ stands for the positive square
roots of the eigenvalues of $m^\dag m$. $\lambda$ should be
thought of as a radial coordinate, i.e. $\lambda\in[0,\infty)$.

A second way to introduce flux, for $q^+\neq0$, $q^-=0$, is to add
a term of the form
\begin{equation}
    S=S_0+iq^+\int(\tr A-\tr B)dt,
\end{equation}
where $A$ and $B$ are the gauge fields of the U($N)_A$ and
U$(N)_B$ gauge groups respectively. This has the effect of
constraining the eigenvalues of $m$ to move in a plane, all with
angular momentum $q^+$. Surprisingly, the reduction to eigenvalues
of $m^\dag m$ gives exactly the potential \eqref{eigpot} with
$M=(q^+)^2-1/4$.

We may try to turn on both $q^+$ and $q^-$ at the same time. It
was shown in \cite{Maldacena:2005he} that this leads again to the
same potential with $M=(|q^+|+|q^-|)^2-1/4$. Thus, we see that
from the point of view of the matrix model, the theory depends on
$|q|\equiv|q^+|+|q^-|$ alone. Due to this and supported by
arguments from the target space theory\footnote{This is also
important for consistency with T-duality to type 0B.}, it was
argued in \cite{Maldacena:2005he} that physics depends only on
$|q|$. This point will be important in the next subsection where we
discuss the two dimensional black hole solution, which requires
turning on both fluxes.
%
\subsection{The Extremal Black Hole Solution}\label{ebh}
The equations of motion \eqref{eomg}-\eqref{eomF} also admit a
solution that is often referred to as the 2d extremal black hole
solution \cite{Gukov:2003yp,Banks:1992xs,Berkovits:2001tg}
\begin{subequations}
\begin{align}
    ds^2&=\left[1+\frac{q^2}8\left(\Phi-\Phi_0-\frac12\right)e^{2\Phi}\right](-dt^2+dx^2),\\
    F^{(+)}&=F^{(-)}=\frac q2dt\wedge dx,\\
    T&=0,
\end{align}\end{subequations}
where $\Phi_0=-\log\frac q4$ and $\Phi$ is given implicitly by the
ODE
\begin{equation}
     \frac1{2}\dif{\Phi}{x}=1+\frac{q^2}8\left(\Phi-\Phi_0-\frac12\right)e^{2\Phi}.
\end{equation}
The boundary conditions are set such that at the asymptotic region
$x\ra-\infty$ the solution approaches the linear dilaton solution.
In the $x\ra+\infty$ region the solution becomes AdS$_2$ with string
coupling $g_s=4/q$ :
\begin{align}\label{ads2met}
    ds^2&\ra\frac1{8x^2}(-dt^2+dx^2),&F^{(+)}&=F^{(-)}=\frac q2dt\wedge dx,&\Phi&\ra
    \Phi_0.
\end{align}
There are two major problems with this solution. The first is that
the curvature becomes large as $x\ra+\infty$, specifically, at the
AdS$_2$ region of this solution $R=-8$ (in string units), so
higher order corrections to \eqref{type0action} are important and
the solution is invalid there. Note that unlike the linear dilaton
background, this solution is not an exact CFT. The second problem
is that this solution requires turning on both of the 0A vector
fields, which, as we mentioned before, implies that one needs to
add strings to the background and take their effects into account.
Furthermore, the study of the background $(q^+,q^-)=(q^+,0)$ shows
that there is no entropy and no classical absorption. Thus, if the
physics indeed depends only on $|q|$, then a black hole is not
expected to exist.

It was suggested in \cite{Strominger:2003tm} that perhaps, despite
these problems, a solution that interpolates between a linear
dilaton region and an AdS$_2$ region exists for the full theory.
If this is the case, then by the AdS/CFT correspondence
\cite{Maldacena:1997re}, the \ads region of the solution would be
dual to a one dimensional CFT (or conformal quantum mechanics
(CQM)), and it was conjectured in \cite{Strominger:2003tm} that
this CQM is the conformal limit of the 0A matrix model. In the
next sections we will analyze this CQM to see what the properties
of such a solution must be.
%
%
%
\section{The Conformal Limit of the Matrix Model}\label{ads2cqm}
The 0A matrix model eigenvalues move in the potential (see Figure 1)
\begin{figure}[h]
\begin{center}
\hspace{-1cm}%
\includegraphics[height=9.0cm]{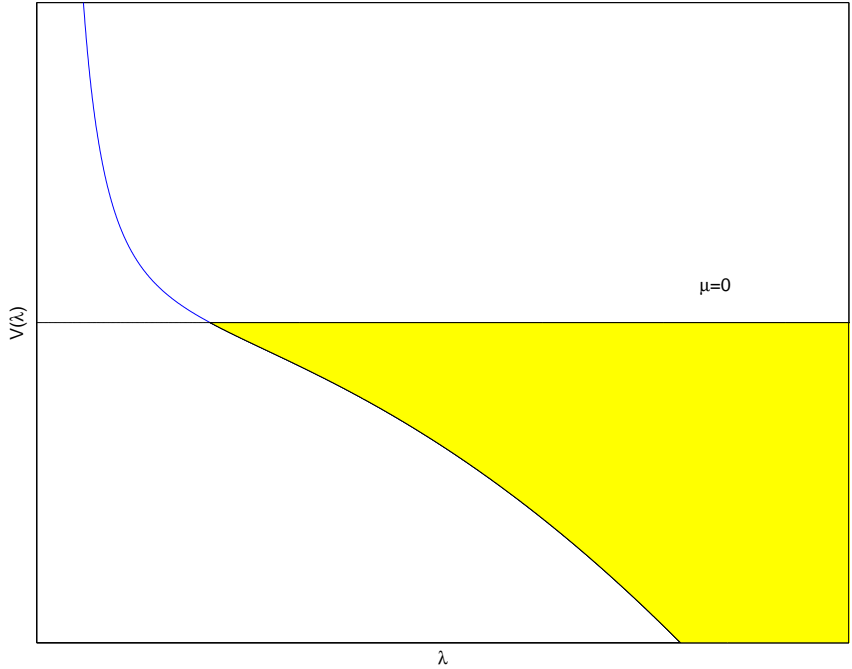}%
\hspace{0.5cm}%
\caption{The potential for the eigenvalues of the matrix model
describing 0A string theory in a background with non-zero flux, with
Fermi level $\mu=0$.}
\end{center}
\end{figure}
\begin{equation}\label{potlam}
    V(\lambda)=-\frac18\lambda^2+\frac M{2\lambda^2}\;.
\end{equation}
At large $\lambda$ the second term is negligible and the dynamics
are as in the original $q=0$ linear dilaton background. At small
$\lambda$ one can ignore the $\lambda^2$ term and remain with the
action
\begin{equation}\label{cqmlag}
    S=\frac12\int dt\left(\dot\lambda^2-\frac M{\lambda^2}\right).
\end{equation}
This action was studied in detail in \cite{deAlfaro:1976je} as the
simplest example of a (nontrivial) conformal field theory in one
dimensional space (only time). It is invariant under the
SL(2,$\Rf$) group of transformations given by
\begin{align}
    &t\ra t'=\frac{at+b}{ct+d},&&
    \lambda(t)\ra\lambda'(t')=(ct+d)^{-1}\lambda(t),
\end{align}
with $ad-bc=1$. The generators of these transformations are
\begin{align}
    H&=\frac12\dot\lambda^2+\frac{M}{2\lambda^2},
    &K&=\frac12\lambda^2,
    &D&=-\frac14(\lambda\dot\lambda+\dot\lambda\lambda).
\end{align}
Note that the AdS/CFT correspondence maps $H$ to time evolution in
the Poincar\'{e} time coordinate of AdS$_2$. It is important to
emphasize that the \ads vacuum is supposed to be mapped to the
solution with $\mu=0$, since a finite Fermi sea would break
conformal invariance. The relation to SO(2,1) is made evident by
taking the linear combinations
\begin{align}
    L_{01}&=S=\frac12\left(\frac1rK-rH\right),
    &L_{02}&=D,
    &L_{12}&=R=\frac12\left(\frac1rK+rH\right).
\end{align}
(for any constant $r$). Here $L_{\mu\nu}$ are rotations in the
$\mu\nu$ plane, thus, the operator $R$ is compact. It should also
be noticed that the three operators are related by a constraint in
this system
\begin{equation}
    \frac12(HK+KH)-D^2=\frac M4-\frac3{16}=\frac {q^2-1}4
\end{equation}
(this is why one seems to have three constants of motion in a two
dimensional phase space).

The essential observation of \cite{deAlfaro:1976je} is that by a
reparametrization of time and field,
\begin{align}
    d\tau&=\frac{dt}{u+vt+wt^2},
    &\tilde\lambda(\tau)&=\frac{\lambda(t)}{\sqrt{u+vt+wt^2}},
\end{align}
one may transform the action to a different action where the new
corresponding Hamiltonian ($\tau$-translation operator) is
$G=uH+vD+wK$. Specifically, by choosing $(u,v,w)=(r,0,r^{-1})$ one
gets the transformation
\begin{align}
    \tilde\lambda(\tau)&=\frac{\sqrt r}{\sqrt{r^2+t^2}}\lambda(t),
    &\tau&=\arctan(t/r).
\end{align}
The transformed action is
\begin{equation}\label{Raction}
    S=\frac12\int d\tau\left[(\dd_\tau\tilde\lambda)^2-
    \tilde\lambda^2-\frac{M}{\tilde\lambda^2}\right],
\end{equation}
whose Hamiltonian is the compact operator $R$. The spectrum of $R$
is found using standard algebraic methods to be\footnote{More
accurately, $r_0$ may also take the value $(1-\sqrt{M+1/4})/2$ if
$M=0$. However, $M=0$ is not relevant for our study since it means
that $q=\pm1/2$, and $q$ is quantized in integer units.}
\cite{deAlfaro:1976je}
\begin{align}\label{Rspec}
    r_n&=r_0+n, &r_0&=\frac12(1+\sqrt{M+1/4})
    =\frac{1+|q|}2\,.
\end{align}
The relations of the $R$ eigenstates to the $H$ eigenstates is
also given in \cite{deAlfaro:1976je}, as well as general methods of
computing transition matrix elements exactly. Obviously, the
spectrum of $H$ is continuous and $D$-transformations rescale the
eigenvalues of $H$.

As usual in the AdS/CFT correspondence, time evolution by $R$ is
supposed to be dual to time evolution of 0A string theory on
AdS$_2$ in global coordinates \cite{Strominger:2003tm}. The SL(2,$\Rf$)
generators in the language of the new  time parameter generate the
transformations
\begin{subequations}
\begin{align}
    &D &\delta\tau&=2\sin\tau &\delta\lambda&=\lambda\\
    &H &\delta\tau&=\frac1r(1+\cos\tau)&\delta\lambda&=0\\
    &K &\delta\tau&=r(1-\cos\tau)&\delta\lambda&=-r\tan(\tau/2)\lambda
\end{align}
\end{subequations}
Indeed, this corresponds to the generators of the \ads isometries
near the boundary \eqref{iso-ads}. A more complete analysis of the
relation between the isometries and different parametrizations of
\ads is given in \cite{Ho:2004qp}.

The original matrix model gives $N$ non-interacting fermions
moving in the full potential \eqref{eigpot}. After taking the CQM
limit, we expect some number of these fermions to ``live" in the
small $\lambda$ region, leading in the large $N$ limit to a second
quantized version of \eqref{cqmlag}. Note that the gauge-invariant
operators in the matrix model are given by
\begin{equation}
    \Lambda_{n}=\tr\left[(m^\dag m)^n\right]=\sum_{i=1}^N\lambda_i^{2n}
\label{gaugeinvops}
\end{equation}
and in the CQM their mass dimension is $(-n)$ (recall that
$\lambda_i$ are the positive roots of the eigenvalues of $m^\dag
m$).
%
%
\section{The \ads Dual of the CQM}\label{strcqm}
We have seen that the spectrum of the operator $R$ in the CQM is
$r_n=r_0+n$ (see \eqref{Rspec}). Thus, since the eigenvalues are
fermionic, the spectrum of $R$ in the second quantized CQM is given
by stating for each level, $r_n$, whether it is occupied or vacant.
We would like to identify this with the spectrum of some global \ads
solution of type 0A string theory, which would be a corrected
version of \eqref{ads2met}. In fact, this is precisely the same as
as the spectrum of a single free fermion field on AdS$_2$. Recall
(as we review in the appendix) that the spectrum of a spinor field
of mass $m$ in global \ads (with radius of curvature $r$)
is\footnote{The computation in the appendix is for a field in a
fixed \ads background. Of course, in our case we expect to have a
theory of gravity on AdS$_2$, but, as in other two dimensional
backgrounds, the graviton-dilaton sector has no physical excitations
and including it leads to the same results.}
\begin{equation}
    E_n=\frac12+|mr|+n.
\end{equation}
The spectra match if we take a spinor on \ads with mass
$mr=|q|/2$. We suggest (based on the origin of the eigenvalues)
that the excitations of this spinor field are brane-anti-brane
pairs. Since these states account for the full spectrum of the
CQM, we suggest that all other fields (in particular the tachyon)
have no physical excitations in AdS$_2$. We have reached this
conclusion by analyzing the spectrum in global coordinates, but of
course it should apply to the Poincar\'e coordinates of \ads as well.

We expect that the matrix model with the original potential
\eqref{eigpot} and with Fermi level $\mu=0$ should correspond to a
flux background of 0A which interpolates between a weakly coupled
linear dilaton region and a Poincar\'e patch of \ads (as in the
extremal black hole solutions of section \ref{ebh}).
We can check if the absence of the tachyon field in the \ads
region is consistent with this expectation. In the linear dilaton
region, tachyon excitations are mapped to excitations of the
surface of the Fermi sea in the matrix model. Before taking the
``near horizon" limit  the classical trajectory of a fermion
moving in the potential \eqref{eigpot} with energy
$E=(\alpha')^{-1/2}\eps$ has a turning point at
\begin{equation}
    \lambda_{\x{max}}=(\alpha')^{1/4}\sqrt{\sqrt{2M+\eps^2}-\eps}
\end{equation}
(where we have reinstalled $\alpha'$). Since the quadratic term in
\eqref{eigpot} has a coefficient $1/(4\alpha')$, the conformal
limit is achieved by taking $\alpha'\ra\infty$. We wish to
consider what happens to a state in the matrix model
\eqref{eigpot} corresponding to a tachyon excitation (with a
finite energy in string units) when we take this limit, so we keep
$\eps$ fixed. Clearly, the limit drives the turning point to
infinity. Namely, a finite excitation of the surface of the Fermi
sea in the asymptotic region of the original matrix model does not
penetrate into the region that we are interested in. On the other
hand, fermions with very high excitation energies can penetrate
into the CQM region, and we identify them with the fermionic
excitations on \ads discussed above.

This result suggests that the string dual of this \ads background,
expected to be a strongly coupled theory on the worldsheet, has the
property that all closed string states in the theory are
non-physical (except for, perhaps, discrete states at special
momenta). We also predict that the brane-anti-brane excitations of
this theory have masses proportional to the RR flux. This suggests
that perhaps the string coupling of the dual is $g_s\sim 1/q$ as in
\eqref{ads2met} \cite{Gukov:2003yp,Strominger:2003tm}, but it is not
clear how to define the string coupling in the absence of string
states.

Naively one may have expected that the theory on \ads should contain
bosonic fields which are dual to the gauge-invariant operators
\eqref{gaugeinvops} of the matrix model, as usual in the AdS/CFT
correspondence. At first sight this seems to be inconsistent with
the fact that we find no bosonic fields in the bulk. Presumably, the
operators \eqref{gaugeinvops} are mapped to complicated combinations
of the fermion field we found.

One of the general mysteries associated with \ads backgrounds in
string theory is the fact that they can fragment into multiple
copies of \ads (related to the possibility for extremal black
holes to split) \cite{Maldacena:1998uz}. Here we find no sign of
this phenomenon. Presumably this is related to the fact that there
are no transverse directions for the D-branes to be separated in.
%
%
\section{Remarks on Bosonization}\label{boso}

Even though we found that the spectrum of the CQM can be identified
with that of a free fermion field on AdS$_2$, it is interesting to ask if
there could also be an alternative bosonic description of the same
theory, which could perhaps be interpreted as a closed string dual
description. The context in which it seems most likely that such a
description would exist is when we look at the CQM \eqref{cqmlag} with
time evolution by $H$, and turn on a finite positive Fermi level
$\mu$. This will clearly break the SL(2,$\Rf$) conformal symmetry (and
hence the spacetime isometry on the string side), so it should no
longer be dual to an \ads background (note that such a state has
infinite energy, so perhaps the corresponding background is not even
asymptotically AdS$_2$). In such a configuration there are excitations
of the surface of the Fermi sea with arbitrarily low energies, and
these could perhaps be mapped to the tachyon field, as was the case in
\eqref{eigpot} before taking the ``near horizon" limit. Thus, it is
interesting to search for a possible alternative description of this
state (except for filling the Fermi sea of the brane-anti-brane
excitations).

A method that has proven to be very fruitful for studying the matrix
model in the past is that of the ``collective field formalism". This
method of bosonization, achieved by studying the dynamics of the
Fermi liquid in phase space, has led to the computation of
scattering amplitudes and other important quantities in the matrix
model (e.g. \cite{Polchinski:1994mb,Demeterfi:1993cm,
Demeterfi:1993sj}). However, there is an important difference
between the CQM and the standard linear dilaton matrix model: in the
CQM for momentum $p$ and $t\ra\pm\infty$ we have $\lambda\sim p\,t$,
while in the linear dilaton case we have $p\sim\mp\lambda$. Suppose
that at some finite time we construct a small pulse perturbing the
shape of the Fermi sea. The relation $\lambda\sim p\,t$ means that
after (and before) a finite amount of time the pulse will ``break
up", or more explicitly, the pulse will no longer admit a
description in terms of the upper and lower surfaces of the Fermi
liquid\footnote{The collective field formalism breaks down when the
Fermi liquid in phase space can no longer be described in terms of
its upper and lower surface $p_+(\lambda)$ and $p_-(\lambda)$.}. We
can follow through the steps of the collective field formalism in
order to analyze propagation of pulses along short time intervals or
calculate some other quantities\footnote{For examples, one can
calculate the free energy, which turns out to be $E_0=(4/3)\mu^2L$,
where $L$ is an IR-cutoff.}, but the elementary excitations of the
bosonized version will not be asymptotic states.

This asymptotic behavior of the classical solutions is, of course, a
consequence of the fact that the potential is constant  as
$\lambda\ra+\infty$. One may thus seek other methods of bosonization
that are suitable for systems with this property. However,
the resulting bosonic systems always seem to have non-local
interactions (for instance, this happens in the method presented in
\cite{Dhar:2005fg}). We have not been able to find a bosonic theory
with local interactions, which could be interpreted as a bosonic field
on some spacetime dual to the CQM with finite $\mu$. It would be
interesting to investigate this further.

%
\section*{Acknowledgements}
We would like to thank M. Berkooz, D. Reichmann, A. Strominger and
T. Takayanagi for useful discussions. This work was supported in
part by the Israel-U.S. Binational Science Foundation, by the
Israel Science Foundation (grant number 1399/04), by the
Braun-Roger-Siegl foundation, by the European network
HPRN-CT-2000-00122, by a grant from the G.I.F., the German-Israeli
Foundation for Scientific Research and Development, by Minerva,
and by the Einstein center for theoretical physics.
%
\appendix
\renewcommand{\thesubsection}{\Alph{section}.\arabic{subsection}}
\section{Spinor Fields on AdS$_2$}\label{qftonads2}
\subsection{Definitions of AdS$_2$}
The metric on AdS$_2$ may be written in so called global
coordinates as
\begin{equation}
    ds^2=\frac{r^2}{\cos^{2}\theta}(-d\tau^2+d\theta^2),
\end{equation}
where $\tau\in\Rf$, $\theta\in[-\pi/2,\pi/2]$, and $r$ is the AdS
radius of curvature. This is (conformally) an infinite strip, the
boundaries at $\theta=\pm\pi/2$ being each a line. Another set of
coordinates is the Poincar\'e coordinates ($t\in\Rf$ and
$x\in\Rf^+$), where the line element is given by
\begin{equation}
    ds^2=\frac{r^2}{x^2}(-dt^2+dx^2).
\end{equation}
The relation between the coordinate sets is
\begin{align}
   &x=\frac{r\cos\theta}{\cos\tau-\sin\theta},&&
    t=\frac{r\sin\tau}{\cos\tau-\sin\theta},
\end{align}
or
\begin{align}
    &\tan\tau=\frac{2rt}{x^2-t^2+r^2},
   &&\tan\theta=\frac{t^2-x^2+r^2}{2rx}.
\end{align}
The generators of isometries in the Poincar\'e coordinates are
\begin{align}
    H=i\dd_t, && D=i(t\dd_t+x\dd_x), &&K=i((t^2+x^2)\dd_t+2tx\dd_x),
\end{align}
and in global coordinates
\begin{align}
    rH&=i[(1-\cos\tau\sin\theta)\dd_\tau+\sin\tau\cos\theta\dd_\theta],\\
    K/r&=i[(1+\cos\tau\sin\theta)\dd_\tau-\sin\tau\cos\theta\dd_\theta],\\
    D&=i[-\sin\tau\sin\theta\dd_\tau+\cos\tau\cos\theta\dd_\theta].
\end{align}
Near the boundaries these become
\begin{align}\label{iso-ads}
    rH&=i(1\mp\cos\tau)\dd_\tau,
    &K/r&=i(1\pm\cos\tau)\dd_\tau,  &D&=\mp i\sin\tau\dd_\tau.
\end{align}
%
\subsection{Quantization of Spinor Fields on \ads in Global Coordinates}

Consider a spinor field on AdS$_2$ (in global coordinates). In this
section we work with signature $(+,-)$. The vierbein is
\begin{equation}
    V^a_\mu=\frac{r}{\cos\theta}\delta^a_\mu
\end{equation}
($\mu$ is a tensor index and $a$ is an index in the local inertial
frame). The spin connection is
\begin{equation}
    \Gamma_\mu=\frac12\Sigma^{ab}V_a^\nu\nabla_\mu V_{b\nu}
    =\delta_\mu^0\tan\theta\Sigma^{01},
\end{equation}
where
\begin{equation}
    \Sigma^{01}=\tbyt{1/2}00{-1/2}.
\end{equation}
The $\gamma$ matrices are chosen to be
\begin{align}
    &\gamma^0=(r^{-1}\cos\theta)\sigma^1,
    &&\gamma^1=(r^{-1}\cos\theta)i\sigma^2,
\end{align}
so that the Dirac equation, $(i\gamma^\mu\nabla_\mu-m)\psi=0$, may
be written as
\begin{align}\label{dircomp2}
    \tbyt{imr\sec\theta}{\dd_0-\dd_1-\frac12\tan\theta}
    {\dd_0+\dd_1+\frac12\tan\theta}{imr\sec\theta}\psi=0.
\end{align}
Let $\psi(\theta,\tau)=e^{-i\omega\tau}\cos^{1/2}\theta(e^{i\omega
\theta}u(\theta),e^{-i\omega \theta}v(\theta))$. Equation
\eqref{dircomp2} becomes
\begin{equation}
    \dif{}\theta\vt uv=imr\sec\theta\tbyt{0}{-e^{-2i\omega\theta}}
    {e^{2i\omega\theta}}0\vt uv.
\end{equation}
This yields the second order equation
\begin{equation}
    \cos^2\theta
    u''+\cos\theta(2i\omega\cos\theta-\sin\theta)u'-(mr)^2u=0.
\end{equation}
Substituting $z=(1+i\tan\theta)/2$, we have the hypergeometric
equation
\begin{equation}\label{ode}
    z(1-z)u''+(\frac12+\omega-z)u'+(mr)^2u=0.
\end{equation}
The Dirac norm is
\begin{equation}
    (\psi_\omega,\psi_{\omega'})=\int_{-\pi/2}^{\pi/2}d\theta\sqrt{-g}\,
    \psi_\omega^\dag\psi_{\omega'}=
    r^2\int_{-\pi/2}^{\pi/2}\frac{d\theta}{\cos\theta}(u^*u+v^*v),
\end{equation}
so we shall require the solutions of \eqref{ode} to vanish at the
boundaries ($z\ra\infty$). We will show that this requirement
implies that
\begin{equation}\label{quan}
    |\omega|=|mr|+\frac12+n,\qquad n=0,1,2,\cdots
\end{equation}
Equation \eqref{ode} has three (regular) singular points
$\{0,1,\infty\}$. The pairs of exponents at each point are,
respectively,
\begin{equation}
    \{(0,1/2-\omega),(0,1/2+\omega),(mr,-mr)\}.
\end{equation}
When the difference between two exponents at a given point is not
an integer the equation has two power-law solutions at that point
(when this happens we shall call this point ``normal"). When the
difference is an integer there is one power-law and one log
solution (when this happens we shall call the relevant point
``special"). Thus, we see that the proposed result \eqref{quan}
implies that we have three distinct cases: (i) $2mr\not\in\Zf$, in
which case all three points are normal; (ii) $mr\in\Zf$, in which
case all points are special; and (iii) $mr+1/2\in\Zf$, in which
case $z=\infty$ is special and $z=0,1$ are normal.

\subsubsection*{(i) $2mr\not\in\Zf$} In this case there are two power-law
solutions at $z=\infty$. According to the sign of $m$, one of the
solutions diverges at the boundary and the other one is
\begin{equation}\label{odesol}
    u_1(z)=z^{-|mr|}F(|mr|,|mr|+\frac12-\omega,2|mr|+1;z^{-1}).
\end{equation}
As $\theta$ goes from $-\pi/2$ to $\pi/2$, the path drawn by
$z^{-1}(\theta)$ is a unit circle around $z^{-1}=1$. Along this
circle the hypergeometric function has a pole at $z^{-1}=0$ (which
we have taken care of), but may also has a branch cut along
$z^{-1}\in[1,\infty)$. We must make sure that the solution
\eqref{odesol} is continuous (single valued) along the circle. One
option is to take $|mr|+1/2-\omega=-n$ ($n\in\Nf$), so that the
power expansion of the hypergeometric function in \eqref{odesol}
terminates after a finite amount of terms and the function
degenerates into a polynomial. In this case we can choose the
branch-cut to be $(-\infty,0]$, which makes $z^{-|mr|}$ single
valued, and so \eqref{odesol} is single valued.

When $|mr|+1/2-\omega$ is not a non-positive integer the
hypergeometric function will have a branch cut at $[1,\infty)$ and
we must try to cancel this discontinuity with the discontinuity in
$z^{-|mr|}$. For $\omega+1/2\not\in\Zf$, the solution at
$z=\infty$ is related to the solutions at $z=1$ through the
identity
\begin{multline}\label{iden}
    F(|mr|,|mr|+\frac12-\omega,2|mr|+1;z^{-1})=\\=
    \frac{\Gamma(1+2|mr|)\Gamma(1/2+\omega)}{\Gamma(1+|mr|)\Gamma(1/2+|mr|+\omega)}
    F(|mr|,|mr|+\frac12-\omega,\frac12-\omega;1-z^{-1})
    +\\+
    \frac{\Gamma(1+2|mr|)\Gamma(-\frac12-\omega)}{\Gamma(|mr|)\Gamma(|mr|+\frac12-\omega)}
    (1-z^{-1})^{1/2+\omega}F(1+|mr|,|mr|+\frac12+\omega,\frac32+\omega;1-z^{-1}).
\end{multline}
The first of the two terms above has no branch singularity as
$z^{-1}$ circles the point $z^{-1}=1$. The second term has
$\Delta\arg=2\pi(1/2+\omega)$. Since
$\Delta\arg(z^{-|mr|})=2\pi|mr|$ (notice we are moving $z^{-1}$
and not $z$), we must take $\omega=-1/2-|mr|-n$ ($n\in\Nf$), in
which case, the first term above vanishes and the second term
cancels out with the $z^{-|mr|}$. In conclusion, we need
\begin{equation}\label{spectrum}
    |\omega|=|mr|+1/2+n,  \qquad n=0,1,2,\cdots
\end{equation}
When $\omega+1/2\in\Zf$ the identity \eqref{iden} is invalid and
should be replaced by either
\begin{subequations}
\begin{multline}\label{miden}
    F(|mr|,1+|mr|-k,2|mr|+1;z^{-1})=
    -\frac{\Gamma(1+2|mr|)}{\Gamma(|mr|)\Gamma(1+|mr|-k)}
    \times\\\times(z^{-1}-1)^k\log(1-z^{-1})
    \sum_{n=0}^\infty\frac{(|mr|+k)_n(1+|mr|)_n}{n!(n+k)!}(1-z^{-1})^n+G(1-z^{-1})
\end{multline}
or
\begin{multline}\label{miden2}
    F(|mr|,1+|mr|+k,2|mr|+1;z^{-1})=
    -\frac{(-1)^k\Gamma(1+2|mr|)}{\Gamma(|mr|+1)\Gamma(|mr|-k)}
    \times\\\times(z^{-1}-1)^k\log(1-z^{-1})
    \sum_{n=0}^\infty\frac{(|mr|+k+1)_n(|mr|)_n}{n!(n+k)!}(1-z^{-1})^n+G(1-z^{-1}),
\end{multline}\end{subequations}
for $k=0,1,2,\dotsc$, where the function $G(1-z^{-1})$ is some
known function that is single valued as $z^{-1}$ circles
$z^{-1}=1$. Due to the $\log(1-z^{-1})$ these expressions change
by a number that isn't just a phase, so there is no way to cancel
it out with the phase from $z^{-|mr|}$. Thus, for
$\omega+1/2=k\in\Zf$ there is no way to construct a single-valued
solution, and we conclude that \eqref{spectrum} is the only
possibility for $2mr\in\Zf$.

\subsubsection*{(ii) $2mr\in\Zf$ and $m\neq0$}
Let $k/2=|mr|\neq0$ (i.e. $k=1,2...$). In this case the second
solution near $z=\infty$ is
\begin{multline}
    u_2(z)=-(-z)^{-k/2}F(k-\frac12,k-\omega,2k;z^{-1})\log(-z)+(-z)^{-k/2}\sum_{s=0}^\infty c_sz^{-s}+\\
    +(-z)^{k/2}\frac{(-1)^{k-1}k!}{\Gamma(k/2)\Gamma(k/2+1/2-\omega)}
    \sum_{n=0}^{k-1}(-1)^{n}\frac{(k-n-1)!}{n!}\Gamma(n-k/2)\Gamma(n-k/2+1/2-\omega)z^{-n}.
\end{multline}
This solution diverges and should not be considered, so we are
left with the solution \eqref{odesol}. For half-integer $mr$ the
argument of case (i) may be repeated, yielding the same result
\eqref{spectrum}. For integer $mr$, $z^{-mr}$ is single valued, so
we need the hypergeometric function to be single valued as well.
We can do this by taking $\omega=|mr|+1/2+n$ ($n\in\Nf$) as
before, in which case the hypergeometric function degenerates into
a polynomial. Equation \eqref{iden} shows that the function is
never single valued if $\omega+1/2\not\in\Zf$. By \eqref{miden}
and \eqref{miden2}, we can achieve a single valued function for
$\omega+1/2=k=1+|mr|+n$ or $-\omega-1/2=k=|mr|+n$ ($n\in\Nf$), in
which case the multi-valued term with the log vanishes. Together
this reduces to the previous result \eqref{spectrum}.

\subsubsection*{(iii) $mr=0$}
In this case it is most easy to observe that the original equation
\eqref{dircomp2} is solved by
\begin{equation*}
    \psi=e^{-i\omega\tau}\cos^{1/2}\theta\vt{Ae^{i\omega
\theta}}{Be^{-i\omega \theta}}
\end{equation*}
and the two chiral components decouple as expected. These
solutions are only $\delta$-function normalizable, as expected of
massless fields. Again, requiring single-valuedness leads to the
condition \eqref{spectrum}.
%
\subsection{The Dirac Equation in Poincar\'e Coordinates}

We next consider fermions quantized on the Poincar\'e patch, in
order to relate the mass of the fermion with the weight of the
corresponding operator in the CQM. The vierbein is
\begin{equation*}
    V^a_\mu=\frac{r}{x}\delta^a_\mu.
\end{equation*}
The spin connection is
\begin{equation}
    \Gamma_\mu=\frac12\Sigma^{ab}V_a^\nu\nabla_\mu V_{b\nu}
    =\frac1x\delta_\mu^0\tbyt{1/2}00{-1/2}.
\end{equation}
The $\gamma$ matrices are chosen to be
\begin{align}
    &\gamma^0=(x/r)\sigma^1,
    &&\gamma^1=(x/r)i\sigma^2,
\end{align}
so that the Dirac equation, $(i\gamma^\mu\nabla_\mu-m)\psi=0$, may
be written as
\begin{equation}
    \tbyt{imr}{x(\dd_0+\dd_1)-\frac12}
    {x(\dd_0-\dd_1)+\frac12}{imr}\psi=0.
\end{equation}
Let $\psi(x,t)=e^{-i\omega t}x^{1/2}(e^{-i\omega x}u(x),e^{i\omega
x}v(x))$. We have
\begin{equation}
    x\dif{}x\vt uv=\tbyt{0}{imre^{2i\omega x}}
    {-imre^{-2i\omega x}}0\vt uv.
\end{equation}
This yields the second order equation
\begin{equation}
    x^2u''+x(1-2i\omega x)u'-(mr)^2u=0.
\end{equation}
In general, this is solved by
\begin{equation*}
    u=x^{-1/2}e^{i\omega x}(A_+M_{1/2,mr}(2i\omega x)+A_-M_{1/2,-mr}(2i\omega
    x))
\end{equation*}
where $M_{k,m}$ is the Whittaker function. The $A_{\pm}$ terms
behave near $x=0$ as $x^{\pm mr}$.

The boundary condition at infinity should thus be defined by
\begin{equation*}
    \lim_{x\ra0}\psi(t,x)=x^{1/2+|mr|}\psi_0(t)
\end{equation*}
with a finite $\psi_0(t)$, where $\psi_0(t)$ is identified with a
source for an operator $\mathcal O$ in the dual CFT, $\int
dt\psi_0(t)\mathcal{O}(t)$. Therefore, the corresponding operator
$\mathcal O$ has conformal
(mass) dimension $1/2-|mr|$. By relating the spectrum of the CFT
(see \eqref{Rspec}) to the spectrum of the spinor field in
global coordinates, we find in section 4 that the mass of the fermion should
be $|mr|=|q|/2$, so the
operator $\mathcal O$ has mass dimension $(1-|q|)/2$.

%
%
%
\bibliographystyle{jhep}
\bibliography{ads2cqm}
\end{document}